\documentclass[12pt]{article}
\pdfoutput=1
\usepackage[a4paper]{geometry}
\usepackage{jheppub, amsmath,amssymb,amsfonts,amsxtra, mathrsfs, makeidx,graphics,graphicx,amsthm,epsfig,bm,longtable,float, color,tikz,mathtools,xfrac,footnote,rotating, lscape, makecell, environ,mathtools, empheq}
\usepackage{caption}
\usepackage{booktabs}
\usepackage{hyperref}

\usepackage{epsfig,amssymb} 
\usepackage{amsmath}

\usepackage{amscd,color}
\usepackage{amsmath,graphicx}
\usepackage{graphicx}
\usepackage{verbatim}
\usepackage{latexsym}
\usepackage{amsmath,amsfonts,amssymb,amsthm}
\usepackage{amsmath,amsthm}

\def\be{\begin{equation}}
\def\ee{\end{equation}}
\DeclarePairedDelimiter{\abs}{\lvert}{\rvert}

\title{
\begin{center}
A note on 4d $\mathcal{N}=3$ from little string theory
\end{center}}

\author[a]{Antonio Amariti,} 
\author[b]{Gianmarco Formigoni} 
\affiliation[a]{INFN, Sezione di Milano, Via Celoria 16, I-20133 Milano, Italy}
\affiliation[b]{Dipartimento di Fisica, Universit\`a di Milano, \\  Via Celoria 16, I-20133 Milano, Italy}

\emailAdd{antonio.amariti@mi.infn.it,gianmarco.formigoni@gmail.com}
\abstract{
We study little string theory (LST) compactified on $\mathbf{T}^2$, partially breaking 
supersymmetry by a discrete T-duality twist acting on both the 
K\"ahler and  the complex structure of the torus.
This setup gives raise to 4d $\mathcal{N}=3$ models
and it can be performed in both the type IIA and type IIB LSTs.
We comment on the relation with other constructions proposed in the literature.
}

\begin{document}
\maketitle

\section{Introduction}

The study of supersymmetric non-Lagrangian theories has been an intense field of research
in the last decade, after the breakthrough of \cite{Gaiotto}.
It has been shown that they are ubiquitous, and they have played a relevant role in 
the analysis of non-perturbative  phenomena, in their connection with localization, integrability and 
compactification.
An important consequence of the existence of non-Lagrangian theories is that they can 
evade some QFT exact results that follow from the Lagrangian description.

For example it has been recently observed that theories with twelve supercharges
can exist in 4d, but that they must necessarily be non-Lagrangian. 
Indeed such models have not been investigated in the past because they 
can only correspond to strongly coupled isolated fixed points and they
cannot have a perturbative regime. In other words perturbative definitions
of 4d models with $\mathcal{N}=3$ supersymmetry necessarily enhance to $\mathcal{N}=4$ 
 \cite{Weinberg:2000cr}. Nevertheless purely
$\mathcal{N}=3$ theories have been obtained in \cite{Garcia-Etxebarria:2015wns,Aharony:2016kai,Garcia-Etxebarria:2016erx}
(see also  \cite{Bourton:2018jwb,Argyres:2019ngz,Argyres:2019yyb} for alternative constructions
in terms of the discrete gauging of a subgroup of the global symmetry of $\mathcal{N}=4$ SYM.).
These construction required a 
higher dimensional stringy background and a twist in terms of a discrete symmetry
related to the S-duality group. In the stringy inspired constructions the S-duality group 
is associated to other dualities  and symmetries of the gravitational description.
Further checks and generalizations of these proposals have been made in 
\cite{Nishinaka:2016hbw,Imamura:2016abe,Imamura:2016udl,Agarwal:2016rvx,Garcia-Etxebarria:2017ffg,vanMuiden:2017qsh,Cornagliotto:2017dup,Borsten:2018jjm,Arai:2018utu}.

In the constructions of  \cite{Garcia-Etxebarria:2015wns}
the authors considered an F-theory setup, reduced it to 4d $\mathcal{N}=4$
SYM and at the same time they applied an opportune twist  to mod out some of the supercharges.
This twist involved a discrete subgroup of the R-symmetry group combined with 
a twist by the S-duality symmetry of the 4d theory, that had a geometric origin in the higher 
dimensional stringy description.
The two effects did not completely cancel out, giving rise to the possibility of constructing theories with a lower 
amount of supercharges, precisely twelve.

The S-duality twist boils down 
to introduce a non--perturbative object in type IIB string theory, denoted as S-fold in \cite{Dabholkar:2002sy}, 
that generalizes the orientifold projection. 
In s subsequent paper \cite{Garcia-Etxebarria:2016erx} the authors provided an
M-theory construction and related this to an opportune reduction of 6d $\mathcal{N}=(2,0)$.
They considered M-theory on 
$\mathbb{R}^{1,5} \times \mathbb{C} \times \mathbf{T}^3$, by wrapping the M5
branes on a $\mathbf{T}^2$ inside $\mathbf{T}^3$ \cite{Garcia-Etxebarria:2016erx}.
By carefully analyzing the T-duality structure of the theory one can reconstruct 
the twist leading to twelve supercharges in 4d.

In this short note we propose an alternative field theoretical mechanism, 
starting from a 6d non-local field theory, in which gravity 
is completely decoupled, but T-duality remains as a symmetry.
These theories are known as little string theories (LST) 
\cite{Losev:1997hx,Seiberg:1997zk,Berkooz:1997cq} (for review see \cite{Aharony:1999ks,Kutasov:2001uf}) and they correspond to 
6d theories with maximal supersymmetry, either $\mathcal{N}=(1,1)$ or $\mathcal{N}=(2,0)$.

Once these 6d theories are compactified on a 2-torus, the $O(2,2;
\mathbb{Z})$ T-duality group  contains two $SL(2,\mathbb{Z})$ factors.
One of them  acts on the complex structure and the other on the K\"ahler structure of the 2-torus\footnote{The $SL(2,\mathbb{Z})$ transformation on the K\"ahler structure is a TsT transformation
in string theory. Indeed one can first T-dualize on a circle, then perform an $SL(2,\mathbb{Z})$ transformation
on the complex structure and then perform another T-duality. Such TsT corresponds to an $SL(2,\mathbb{Z})$ on the K\"ahler structure \cite{Dhokarh:2008ki}.}.
Depending on the supersymmetry we started with,
 $\mathcal{N}=(1,1)$ or $\mathcal{N}=(2,0)$, 
 one of these two $SL(2,\mathbb{Z})$ symmetries becomes the S-duality group of the 4d theory.
 As we will see, the other $SL(2,\mathbb{Z})$ factor contains a discrete group that has 
 to be considered in the discrete twist (the so called R-twist)  in order  to preserve some supercharges.
This last discrete group is necessary because the little sting theory  
has only an $SO(4)_R$ symmetry group, while the full $SO(6)_R$ 
is manifest only in the limit of vanishing torus. The full discrete symmetry subgroup of the 4d theory is here reconstructed in terms 
of discrete subgroups of $SO(4)$ and of $SL(2,Z)$.
Using this construction we obtain $\mathcal{N}=3$ 4d theories starting from 6d theories with $A_n$ type simply laced gauge groups. 

The note is organized as follows.
In section \ref{sec:LST} we review the basic aspects of LST on $\mathbf{T}^2$, focusing on the geometric origin of the S-duality arising from T-duality of the 6d setup.
In section \ref{sec:LST2} we study the discrete twist that gives origin to the 4d models with 12 supercharges, We discuss both the type IIA and the type IIB constructions.
In section \ref{sec:Mtheory} we compare our construction with the one of \cite{Garcia-Etxebarria:2016erx}.
In section \ref{sec:concl} we conclude discussing further generalizations of our construction.

\section{LST on $\mathbf{T}^2$}
\label{sec:LST}
Little string theories are non-local field theories in 5D and 6D \cite{Losev:1997hx}. 
They can be constructed considering $k$ parallel and overlapping NS 5-branes and then sending the string coupling $g_s \to 0$. Another way to obtain these theories consists of considering $k$ M5-branes with a transverse circle of radius $R$ in the limit $R \to 0$, $M_p \to \infty$ and $R M_p^3 = M_s^2$ constant. After compactification on a $T^d$, LSTs exhibit a $O(d,d;\mathbb{Z})$. 
The type IIA and IIB LSTs are T-dual upon compactification on a circle. 

The compactification of LSTs on a $\mathbf{T}^2$, with radii $r_1$ and $r_2$, gives raise to  $\mathcal{N}=4$ Here we turn on also a B-field with flux given by
\begin{equation}
B=\frac{\theta\alpha'}{r_1r_2}.
\end{equation}
Aa a consequence of the presence of the B-field one finds that the K\"ahler 
structure parameter of the torus is
\begin{equation}
\rho=\frac{ir_1r_2}{\alpha'} + \theta.
\end{equation}
Following \cite{V} and \cite{AKS} we observe that that T-duality on the torus corresponds to the S-duality
of $\mathcal{N}=4$ SYM acting on the holomorphic gauge coupling $\tau_{gauge}=\frac{\theta}{2\pi}+\frac{4\pi i}{g^2_{SYM}}$\footnote{$\theta$ and $g_{SYM}$ are, respectively, a theta angle and the coupling constant of the four dimensional SYM theory.} under the group $SL(2,\mathbb{Z})$. This last  acts as
\begin{equation}
\tau_{gauge} \mapsto \tau'_{gauge}=\frac{a\tau_{gauge}+b}{c\tau_{gauge}+d}
\end{equation}
with the $SL(2,\mathbb{Z})$ matrix of the form $\bigl ( \begin{smallmatrix} a & b \\ c & d \end{smallmatrix} \bigr )$ and $a,b,c,d$ integers such that $ad-bc=1$.

To avoid any confusion, we referred here to the holomorphic gauge coupling
of $\mathcal{N}=4$ SYM as $\tau_{gauge}$ and we  distinguished it from the complex structure to the torus, identified by $\tau$. Furthermore, T-duality on  $S^1 \subset \mathbf{T}^2$  can be used to 
exchange $\tau$ and $\rho$. In this way we will be able to either
associate the K\"ahler (in type IIB) or the complex structure (in type IIA) 
to the holomorphic gauge coupling.

Here we study the type IIB case, where the identification is $\tau_{gauge} = \rho$ and we perform a T-duality on the torus considered above. 
This can be done after fixing the metric and the B-field. 
The metric on the torus can be taken to be
\begin{equation}
g_{\mu\nu}=
\begin {pmatrix}
r_1^2/\alpha' & 0\\
0 & r_2^2/\alpha'
\end {pmatrix}
\end{equation}
where we have rescaled the radii of the torus as
$r_{1,2} \mapsto r_{1,2}/\sqrt{\alpha'}$.
Furthermore, observing that the K\"ahler structure parameter has the  form
$\rho=\rho_1+i\rho_2=B_{21}+i\sqrt{\det g}$ and that the field $B_{\mu\nu}$ is antisymmetric,
we have
\begin{equation}
B_{\mu\nu}=\theta 
\begin {pmatrix}
0 & -1\\
1 & 0
\end {pmatrix}
\end{equation}
At this point one can use the Buscher rules \cite{BUSCHER}. First we apply a T-duality along $r_1$ obtaining\footnote{We used the symbols $\widehat{g}_{\mu\nu}$ and $\widehat{B}_{\mu\nu}$ to refer to quantities after T-duality was performed along the $r_1$ direction.}
\begin{equation}
\widehat{g}_{\mu\nu}=
\begin {pmatrix}
\frac{\alpha'}{r_1^2} & \frac{\theta\alpha'}{r_1^2} \\
\frac{\theta\alpha'}{r_1^2} & \frac{r_1^2r_2^2+\theta^2\alpha'^2}{\alpha'r_1^2}
\end {pmatrix}, \qquad \widehat{B}_{\mu\nu}=0.
\end{equation}
Next we perform the second T-duality in the $r_2$ direction\footnote{The symbols $\widetilde{g}_{\mu\nu}$ and $\widetilde{B}_{\mu\nu}$ indicate quantities after T-duality was performed along the $r_2$ directtion.}. 
We find
\begin{equation}
\widetilde{g}_{\mu\nu}=
\begin {pmatrix}
\frac{\alpha'r_2^2}{r_1^2r_2^2+\theta^2\alpha'^2} & 0 \\
0 & \frac{\alpha'r_1^2}{r_1^2r_2^2+\theta^2\alpha'^2}
\end {pmatrix},
\quad \quad
\widetilde{B}_{\mu\nu}=
\begin {pmatrix}
0 & \frac{\theta\alpha'}{r_1^2r_2^2+\theta^2\alpha'^2} \\
\frac{-\theta\alpha'}{r_1^2r_2^2+\theta^2\alpha'^2} & 0
\end {pmatrix}.
\end{equation}
Thus T-duality on   gives the transformations
\begin{equation}
\label{T-rules}
\frac{r_{1,2}^2}{\alpha'} \mapsto \frac{\alpha'r_{2,1}^2}{r_1^2r_2^2+\theta^2\alpha'^2}, \qquad \theta \mapsto -\frac{\theta\alpha'}{r_1^2r_2^2+\theta^2\alpha'^2}.
\end{equation}
The action of T-duality on $\rho$ is then 
\begin{equation}
 \rho=\frac{ir_1r_2}{\alpha'} + \theta
 \quad
 \mapsto 
  \quad
  \rho'= -\frac{\alpha'}{\theta\alpha'+ir_1r_2}=-\frac{1}{\rho}.
\end{equation}
and it corresponds to the action of  $S \in SL(2,\mathbb{Z})$ on the holomorphic gauge coupling $\tau_{gauge}$.
The action of the generator $T$ is instead associated to a shift of the B-field.
It this way we have recovered the action of $SL(2,\mathbb{Z})$ on the K\"ahler structure of the torus on which we have
compactified the type IIB LST.
This construction can be also extended to the type IIA case. The two descriptions are equivalent because they are
related by T-duality on $S^1$.

\section{$\mathcal{N}=3$ from LST}
\label{sec:LST2}
In this section we construct 4d SCFTs with $\mathcal{N}=3$ supersymmetry starting from LSTs on a torus.
We consider first the $\mathcal{N}=(2,0)$ Type IIA case. In this case there is a global $SO(1,5) \times SO(4)_R$ symmetry and the supercharges are referred as 
$Q_{\alpha A}$ and $\bar{Q}_{\alpha \bar{A}}$ where $\alpha$ is the spinor index for $SO(1,5)$ and $A=2$, $\bar{A}=\bar{2}$ are spinor indices for $SO(4)_R$. The supercharges we are considering correspond to $(4,2)$ and $(4,\bar{2})$\footnote{Here the representations of $SO(1,3)$ are thought as representations of $SO(4)=SU(2) \times SU(2)$. Hence, we denote the representation (2,1) as 2 and the (1,2) as $\bar{2}$.}. Once we compactify on a $\mathbf{T}^2$ we find that the global symmetry becomes\footnote{The full $SO(6)_R$ R-symmetry group is recovered only when the size of the torus vanishes.}
\begin{equation}
SO(1,3) \times SO(4)_R \times U(1)_\tau
\end{equation}
The  supercharges transform with respect to this global symmetry group as
\begin{equation}
(2,2)_1 \oplus (\bar{2},2)_{-1} \oplus (2,\bar{2})_1 \oplus (\bar{2},\bar{2})_{-1}.
\end{equation}

The $U(1)_\tau$ symmetry is identified with the rotation on the torus wrapped by the NS 5-branes. The T-duality group has two $SL(2,\mathbb{Z})$ factors. One of the two factors is associated with $U(1)_\tau$, i.e. the module of the complex structure ${\tau=\frac{g_{12}}{g_{22}}+i\frac{\sqrt{\det g}}{g_{22}}}$, which is playing the role of the holomorphic gauge coupling. The other factor is associated to $U(1)_\rho$, i.e. the module of the K\"ahler structure of the torus ${\rho=B+i\sqrt{\det g}}$. 
In both cases we refer to $g$ as the metric on the 2-torus.

In order to associate a $U(1)$ bundle to an $SL(2,\mathbb{Z})$ bundle, the following procedure can be applied \cite{KW,Bianchi:2011qh}.
 We restrict ourselves to an $\mathcal{N}=2$ sub-algebra of the supersymmetry algebra, which has a pair of right-handed supercharges $Q^i_{\dot{A}}$, $i=1,2$ and a single central charge $Z$. They satisfy

\begin{equation}
\{Q^i_{\dot{A}},Q^j_{\dot{B}}\}=\epsilon_{\dot{A}\dot{B}}\epsilon^{ij}Z.
\end{equation}
Let's look at the action of $SL(2,\mathbb{Z})$ on the supercharges to understand how Z transforms under the S-duality group. The central charge is given by
\begin{equation}
Z=\sqrt{\frac{2}{Im\tau}} \, (\vec{m} \quad \vec{n}) \, \vec{\phi} \, \begin{pmatrix} \tau  \\ 1 \end{pmatrix}.
\end{equation}
For simply-laced gauge groups, the $SL(2,\mathbb{Z})$ group acts as 
\begin{equation}
\label{SL}
\begin{matrix}
Im\tau & \mapsto & \abs{c\tau+d}^{-2} Im\tau \\
(\vec{m} \quad \vec{n}) & \mapsto & (\vec{m} \quad \vec{n}) \, M^{-1} \\
\vec{\phi} & \mapsto & \vec{\phi} \\
\tau & \mapsto & (a\tau+b)/(c\tau+d)
\end{matrix}
\end{equation}
with $M=\bigl ( \begin{smallmatrix} a & b \\ c & d \end{smallmatrix} \bigr )$, $a,b,c,d$ integers and $ad-bc=1$. Moreover, $\vec{\phi}$ is the expectation value of the complex scalar which is the $\mathcal{N}=2$ superpartner of the massless gauge fields and $\vec{n},\vec{m}$ are the electric and magnetic charge respectively. 
Under the transformations (\ref{SL}) 
the central charge transforms as
\begin{equation}
\label{transformationZ}
Z \mapsto \frac{\abs{c\tau+d}}{c\tau+d}Z.
\end{equation}
At this point one can define a $U(1)$ symmetry by requiring that the algebra is invariant under such transformation. Let's work out the details. These symmetries, referred as $U(1)$ chiral rotations in \cite{KW}, 
must act as $\exp(i\hat{\phi})$ on the $\bar{Q}$'s, and as $\exp(-i\hat{\phi})$ on the $Q$'s. Combining this fact with the transformation law of the central charge
(\ref{transformationZ}), one finds
\begin{equation}
\{\exp(-i\hat{\phi})Q,\exp(-i\hat{\phi})Q\} \simeq \frac{\abs{c\tau+d}}{c\tau+d}Z.
\end{equation}
Thus, in order for the algebra to be invariant, we find that
\begin{equation}
\exp(-i\hat{\phi})=\biggl(\frac{\abs{c\tau+d}}{c\tau+d}\biggr)^{\frac{1}{2}}.
\end{equation}
The square root means that the group that acts on the supercharges is a double cover of the duality group $SL(2,\mathbb{Z})$.
The charge $q$ assigned to the supercharges is given by $e^{iq\arg{(c\tau+d)}}$. Therefore we see that $q=\pm1$, i.e. $-1$ for the Q's and $+1$ for the $\bar{Q}$'s.
This additional symmetry, which acts as an accidental outer automorphism on the supercharges, characterizes the spectrum of $\mathcal{N} = 4$ SYM theory and is known in literature as Bonus Symmetry \cite{I}. 
The $\exp{(\mp i \hat{\phi})}$ factor acts on the supercharges that transform under the 4 or $\bar{4}$ of $SO(1,5)$. Once we decompose the supercharges into 2's or $\bar{2}$'s we end up with a $\exp{(\mp i/2 \hat{\phi})}$ factor, where $\hat{\phi}=\arg{(c\tau+d)}$.
The supercharges transform under $SO(1,3) \times SO(4)_R \times U(1)_\tau \times U(1)_\rho$ as
\begin{equation}
(2,2)_{1,-1} \oplus (\bar{2},2)_{-1,1} \oplus (2,\bar{2})_{1,1} \oplus (\bar{2},\bar{2})_{-1,-1}.
\end{equation}
Observe that the $SO(4)_R$ $R$-symmetry combines with $U(1)_{\tau}$ and enhances to the $SO(6)_R$ $R$-symmetry of 4d $\mathcal{N}=4$ SYM. 

We can now perform the discrete twist giving raise to the $\mathcal{N}=3$ theory. 
At a geometric level we consider $(\mathbb{R}^{1,3} \times \mathbf{T}^2_{LST})/\mathbb{Z}_k$
and observe the following identification:
\begin{equation*} \begin{matrix}
&\tau=i& \Longleftrightarrow &g&=&\begin{pmatrix} 0 & -1 \\ 1 & 0 \end{pmatrix}&=&S \\
&\tau=e^{i\pi/3}& \Longleftrightarrow &g&=&\begin{pmatrix} 1 & -1 \\ 1 & 0 \end{pmatrix}&=&TS \\
&\tau=e^{i\pi/3}& \Longleftrightarrow &g&=&\begin{pmatrix} 0 & -1 \\ 1 & -1 \end{pmatrix}&=&(TS)^2
\end{matrix}
\end{equation*}
The values of $\tau=i,e^{i\pi/3}$ are fixed points of the transformation $\tau \to \frac{a\tau+b}{c\tau+d}$. To each value of $\tau$ corresponds a different matrix of the S-duality group $SL(2,\mathbb{Z})$, which can be expressed in terms of the generators S and T. For these specific values of $\tau$, there is an enhancement of the S-duality group $SL(2,\mathbb{Z})$, which becomes a symmetry group of the theory. 
The discrete subgroup associated to $\tau=i$ is $\mathbb{Z}_4$ while $\mathbb{Z}_3,\mathbb{Z}_6$ are associated to $\tau=e^{i\pi/3}$.
Because of these enhancements, it is possible to quotient the $\mathcal{N}=4$ theory by the $\mathbb{Z}_k$ discrete subgroups.\footnote{For the value $k=2$ the discrete quotient corresponds to the usual orientifold projection which preserves 16 supercharges.}
In order to clarify the procedure, we observe that the $\mathbb{Z}_k$ twist is generated by the following factors.
\begin{itemize}
\item The $\mathbb{Z}_k^R$ factor is a rotation generated by the matrix $R_k=\bigl ( \begin{smallmatrix} \hat{R}_k^{-1} & 0 \\ 0 & \hat{R}_k \end{smallmatrix} \bigr )$  of the  $SO(4)_R$ R-symmetry group.
\item The $\mathbb{Z}_k^{\tau}$ factor is associated  to rotation on the torus wrapped by NS 5-branes.
\item The $\mathbb{Z}_k^{\rho}$ factor is associated to the bonus symmetry group.
\item Each of these elements acs as a $e^{\pm i\pi/k}$ factor on the supercharges.
\end{itemize}
The  $\mathbb{Z}_k^{\tau,\rho}$ factors descend from the torus wrapped by the NS 5-branes. Indeed by following the procedure explained above we associate 
 a $U(1)_{\tau,\rho}$ bundle to the discrete $SL(2,\mathbb{Z})_{\tau,\rho}$ groups. 
By acting explicitly on the supercharges with the $\mathbb{Z}_k^{\tau,\rho} \subset U(1)_{\tau,\rho}$  discrete symmetries, 
we can see that some supercharges  are left invariant while some others are modded out
 \footnote{States of the representation 2 of $SO(4)$ are denoted as $(++),(--)$. In a similar fashion states of the representation $\bar{2}$ are denoted as $(+-), (-+)$.}:
\begin{align*}
&(2,++)_{1,-1}& \quad &\mapsto& \quad &(2,++)_{1,-1}& \\
&(2,--)_{1,-1}& \quad &\mapsto& \quad &(2,--)_{1,-1}&  \\
&(\bar{2},++)_{-1,1}& \quad &\mapsto& \quad &(\bar{2},++)_{-1,1}&  \\
&(\bar{2},--)_{-1,1}& \quad &\mapsto& \quad &(\bar{2},--)_{-1,1}&  \\
&(2,+-)_{1,1}& \quad &\mapsto& \quad &(2,+-)_{1,1}& \\
&\color{red}{(2,-+)_{1,1}}& \quad &\mapsto& \quad \color{red}{e^{4\pi i/k}} &\color{red}{(2,-+)_{1,1}}& \\
&\color{red}{(\bar{2},+-)_{-1,-1}}& \quad &\mapsto& \quad \color{red}{e^{-4\pi i/k}} &\color{red}{(\bar{2},+-)_{-1,-1}}& \\
&(\bar{2},-+)_{-1,-1}& \quad &\mapsto& \quad &(\bar{2},-+)_{-1,-1}& 
\end{align*}
From the above expression one can see that only 12 out of the original 16 supercharges survived the projection, giving 
 raise to an $\mathcal{N}=3$ theory.

\subsection{Type IIB Case}

In this case the supercharges are $Q_{\alpha A}$ and $\bar{Q}_{\bar{\alpha}\bar{A}}$ corresponding to $(4,2)$ and $(\bar{4},\bar{2})$. As in the previous case, after compactification on a torus we have the following supercharges
\begin{equation}
(2,2)_1 \oplus (\bar{2},2)_{-1} \oplus (2,\bar{2})_{-1} \oplus (\bar{2},\bar{2})_1.
\end{equation}
where the charge refers to the $U(1)_\rho$ symmetry in this case. Here the $SL(2,\mathbb{Z})_\rho$ is playing the role of S-duality because the module of the K\"ahler structure corresponds to the holomorphic gauge coupling of the 4d theory. We know that in this case there is also an $SL(2,\mathbb{Z})_\tau$,  where that this factor descends from the $O(2,2;\mathbb{Z})$ T-duality group. As a symmetry it corresponds to the maximal compact subgroup $U(1)_\tau$ of $SL(2,\mathbb{R})$. To find how the supercharges transform under $SO(1,3) \times SO(4)_R \times U(1)_\rho \times U(1)_\tau$, one can follow the procedure outlined in the previous section. 
An alternative, bu equivalent derivation consists of exchanging the $U(1)$ charges discussed in the previous type IIA case. 
This is because the two theories, considered on $S^1 \subset \mathbf{T}^2$  are T-dual, and T-duality exchanges $\tau \leftrightarrow \rho$. With this procedure 
one would find the same result up to an overall minus sign, which can be thought as a parity transformation on the supercharges.
 In this case the supercharges are
\begin{equation}
(2,2)_{1,-1} \oplus (\bar{2},2)_{-1,1} \oplus (2,\bar{2})_{-1,-1} \oplus (\bar{2},\bar{2})_{1,1}.
\end{equation}
The discrete twist by $\mathbb{Z}_k$ is  done by combining the discrete subgroups arising from 
$U(1)_\tau$, $U(1)_\rho$ and $SO(4)_R$ as was done in the section above.
The final result is still that there are 12 supercharges left invariant by the projection.

\section{Relations with the literature}
\label{sec:Mtheory}

So far we have obtained a 4d $\mathcal{N}=3$ SCFT
in terms of LST compactified on $\mathbf{T}^2$.
In this section we compare our construction with the others appeared in the literature.
More precisely we compare our construction with the non-geometric one of \cite{Garcia-Etxebarria:2016erx}  
in terms of M5 branes wrapping a $\mathbf{T}^2\subset \mathbf{T}^3$
\footnote{
As we discussed in the introduction, a previous attempt was proposed in \cite{Garcia-Etxebarria:2015wns},
where the authors obtained a geometric construction of 4d $\mathcal{N}=3$ models,
in terms of  F-theory at terminal singularity. Here we skip its review because  it  does not play a prominent role in our analysis.
}.

\begin{figure}
\begin{center}
\begin{tabular}{cc}
\includegraphics[width=6cm]{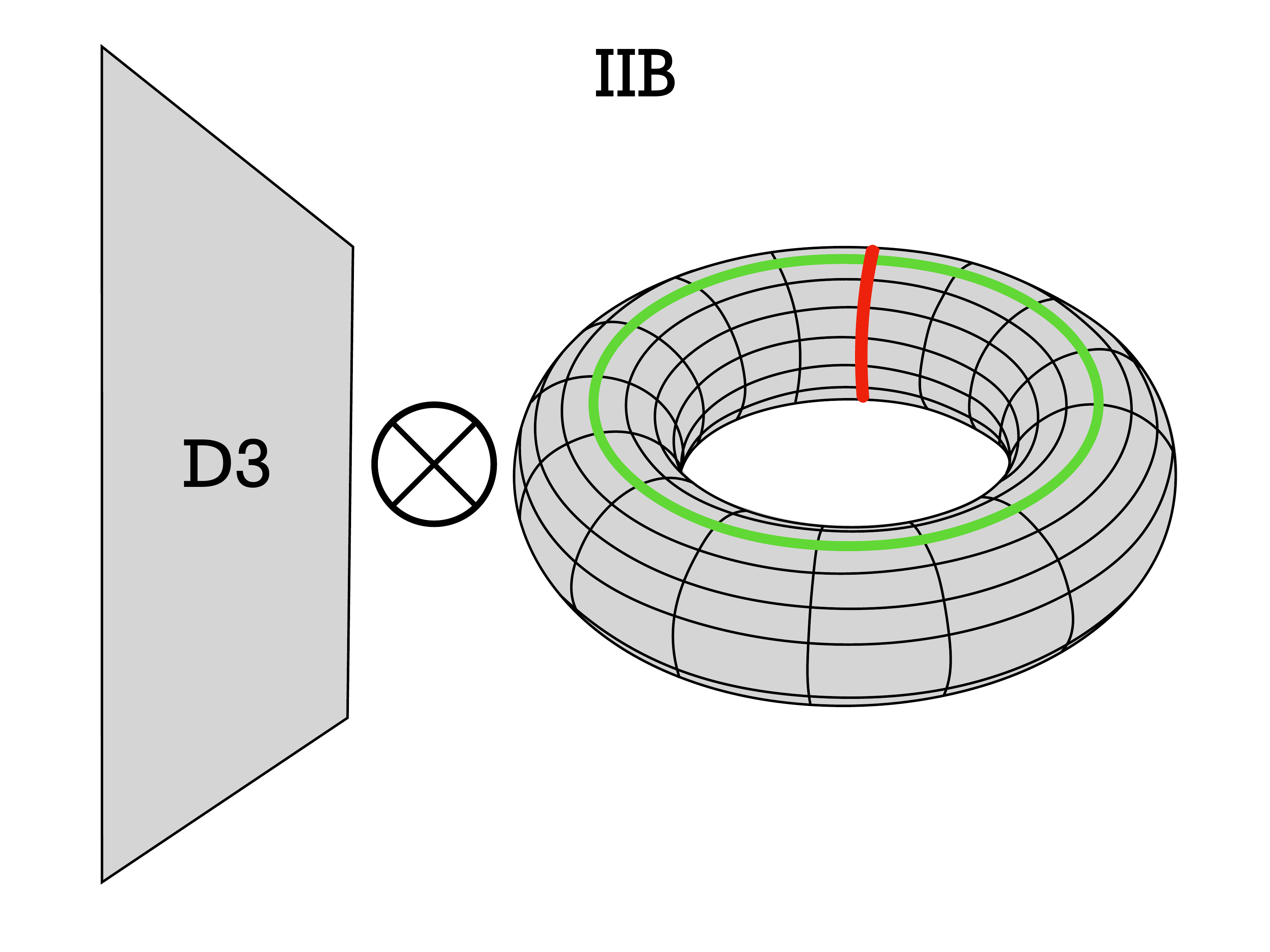}
&
\includegraphics[width=6cm]{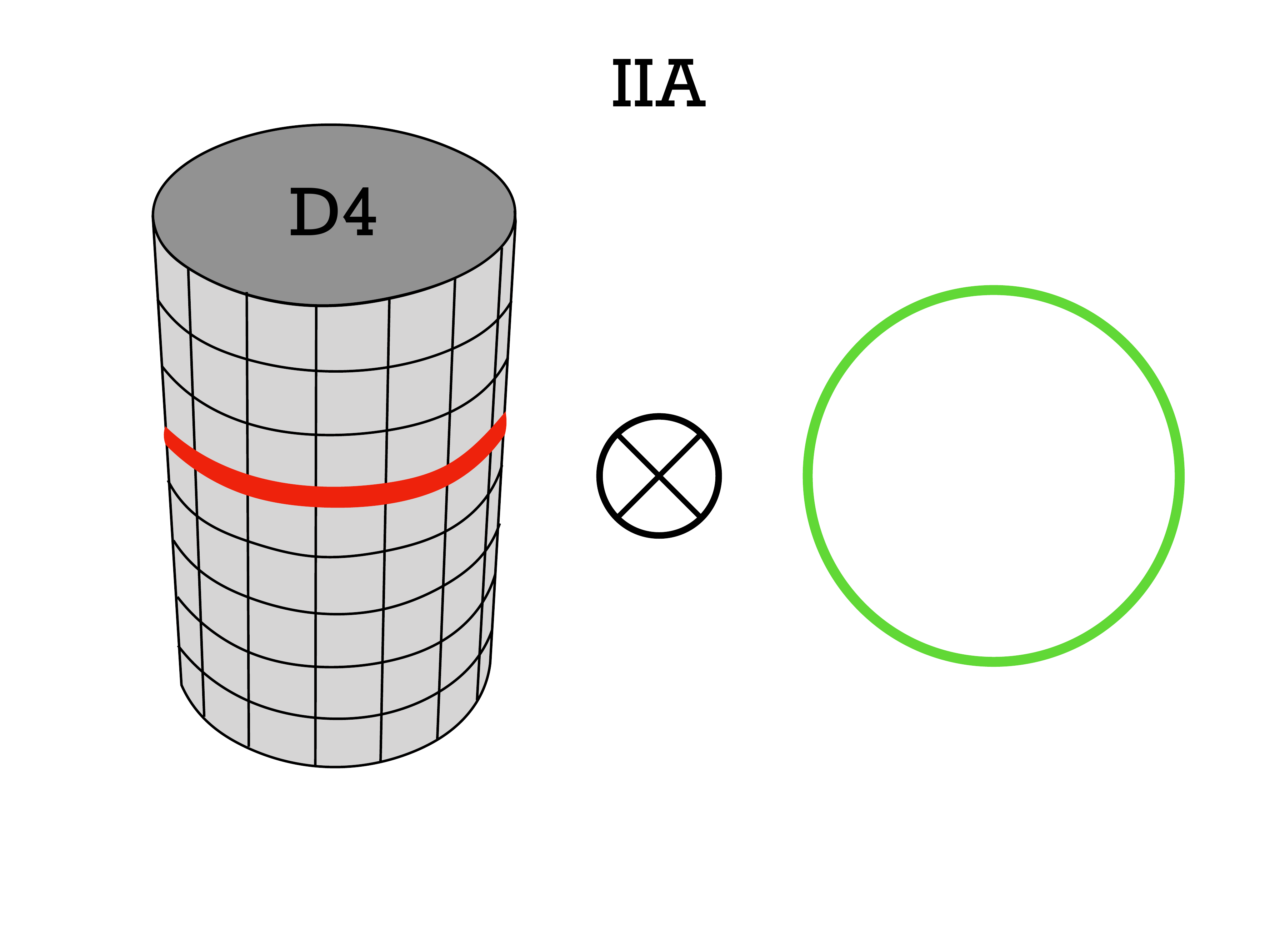}
\\
\includegraphics[width=6cm]{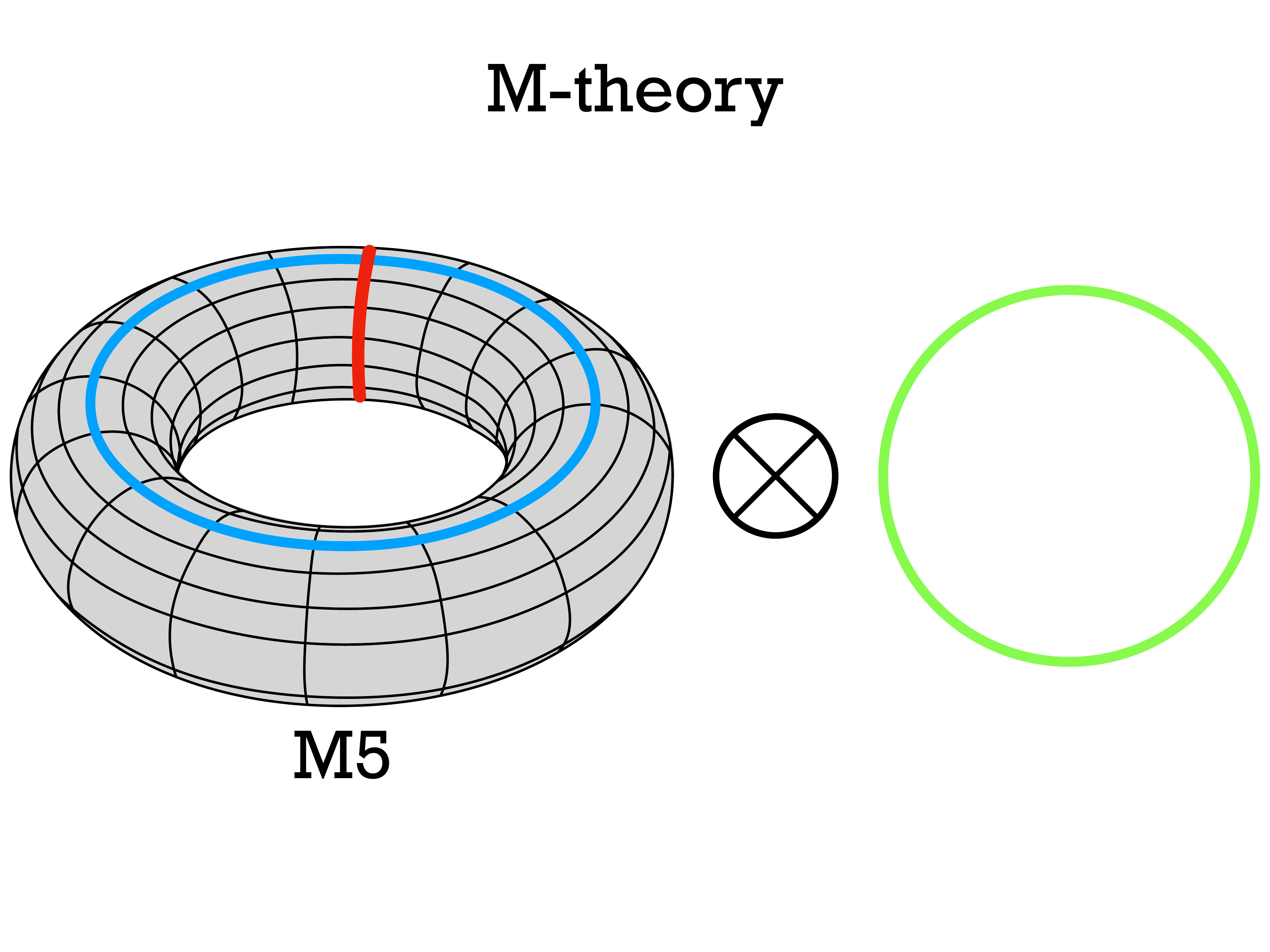}
&
\includegraphics[width=6cm]{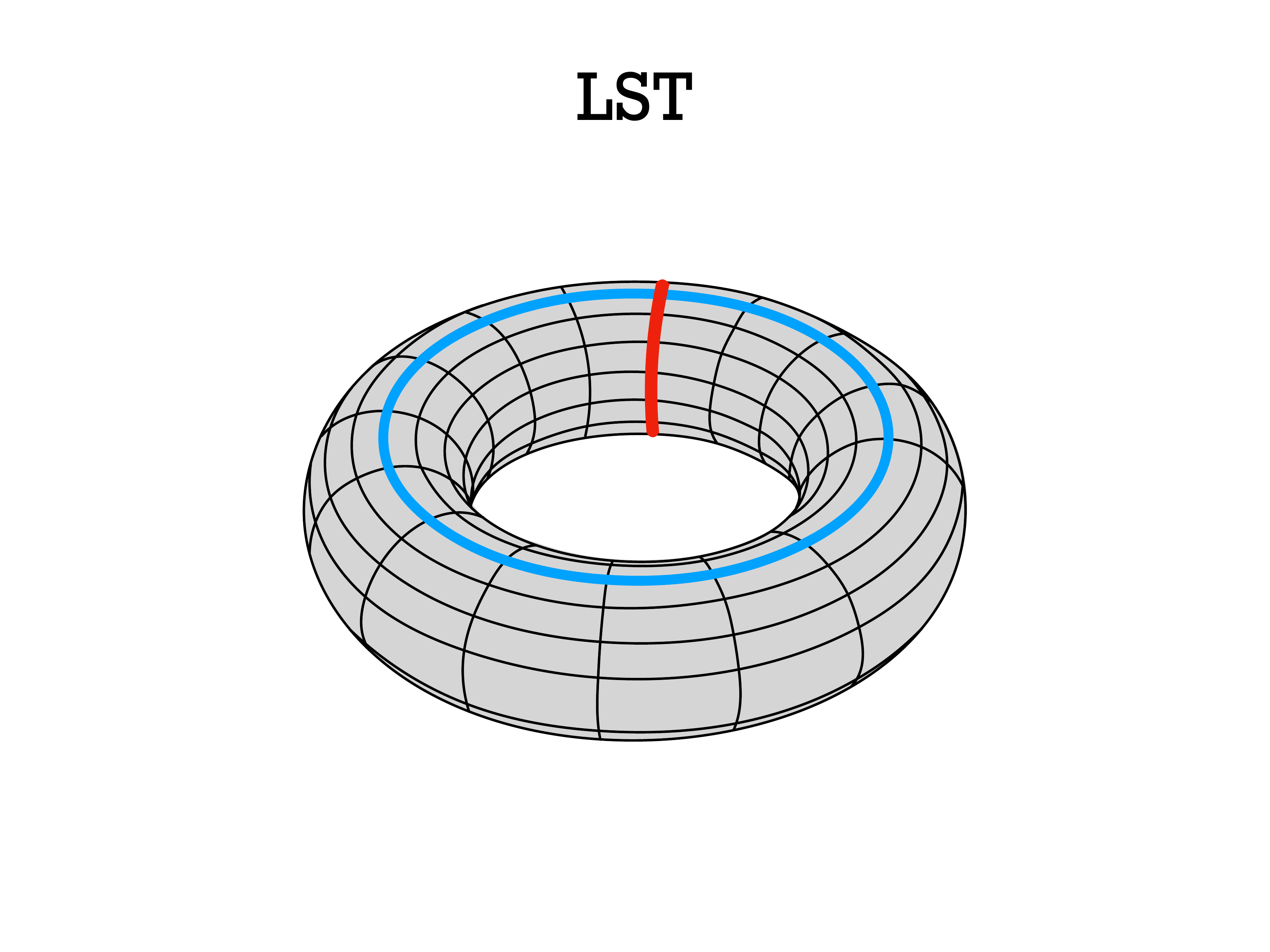}
\end{tabular}
\caption{Representation of the various constructions discussed in   \cite{Garcia-Etxebarria:2016erx} giving raise to 
an $\mathcal{N}=3$ model through the S-duality twist.
The first figure represents the type IIB construction: there is a D3 brane and the transverse space 
is $\mathbb{C}^2 \times \mathbf{T}^2$.
This construction is related through T-duality to the second figure, that represents 
a D4 brane on $\mathbb{R}^{1,3} \times S^1$, with a transverse $\mathbb{C}^2 \times S^1$.
This can be uplifted to M-theory, as shown in the third figure. 
The last figure represents our construction that 
can be obtained from the M-theory one by reducing along the \emph{red} circle.
The last picture represents an NS5 brane wrapping a $\mathbf{T}^2$, and it is evident from the figure that
all the informations necessary for the S-duality twist are in the 6d picture, i.e. in the 6d non-local 
field theory denoted as LST.}
\label{fig}
\end{center}
\end{figure}

Let us review the construction of \cite{Garcia-Etxebarria:2016erx}.
The S-fold projection acts on a 4d theory corresponding to a stack of D3 branes in type IIB.
The discrete twist requires a $\mathbf{T}^2$ in the transverse geometry, as shown in Figure \ref{fig}.
The torus breaks $SO(6)_R \rightarrow SO(4)_R$, and the full R-symmetry group is recovered 
only in the decompactification limit. This picture can be T-dualized to type IIA and lift to M-theory 
as in Figure \ref{fig}.

In this case we have M-theory on $\mathbb{R}^{1,3} \times \mathbf{T}^3 \times \mathbb{C}^2$
with a stack of M5  branes wrapping a $\mathbf{T}^2 = {\color{red} S^1} \times {\color{blue}S^1}$ inside 
$\mathbf{T}^3 = {\color{red} S^1} \times {\color{green}S^1} \times {\color{blue} S^1}$.
The K\"ahler structure parameter is given by $\rho = \int_{\mathbf{T}^3} C + i \sqrt {\det{G}}$ 
and it plays the role of the holomorphic gauge coupling of the $\mathcal{N}=4$ theory.
We denoted with $C$ the M-theory three form and $G$ the metric on $\mathbf{T}^3$.
The M-theory duality group in this background is $SL(3,\mathbb{Z})\times SL(2,\mathbb{Z})_{\rho}$.
The authors showed that if the radii are constrained as 
$r_{\color{red}S^1}=r_{\color{blue}S^1}=r_{\color{green}S^1}^{-1/2}=r $
(or $r_{\color{red}S^1}=\frac{2 \sqrt{3} }{3} r_{\color{blue}S^1}=r_{\color{green}S^1}^{-1/2}=r $)
there is a discrete symmetry enhancement giving raise to the S-duality twist.

Here we observe that our construction can be obtained from the one of 
\cite{Garcia-Etxebarria:2016erx} by shrinking the M-theory circle $r_{\color{green}S^1}$.
This limit correspond to $r \rightarrow 0$ while keeping either 
$\frac{r_{\color{red}S^1}}{r_{\color{blue}S^1}}=1$ or
$\frac{r_{\color{red}S^1}}{r_{\color{blue}S^1}}=\frac{\sqrt{3}}{2}$.
We are left with NS5 branes 
wrapping s $\mathbf{T}^2$ , where we kept the same colors for the radii in order to make the relation explicit.
In this case the $SL(3,\mathbb{Z})\times SL(2,\mathbb{Z})_{\rho}$ duality group of 
 M-theory becomes the usual T-duality group that survives the decoupling of the gravitational degrees of freedom 
 in LST.
 The M-theory three form $C_{\mu \nu\rho}$ becomes the two form $B_{\mu\nu}$ in LST while the R-symmetry 
 group remains $SO(4)_R$.

\section{Conclusions}
\label{sec:concl}
In this note we studied 4d $\mathcal{N}=3$ SCFT from LST on a $\mathbf{T}^2$,  obtained by a twist of 
a discrete symmetry subgroup of the full T-duality group.
This is an example of a purely field theoretical construction, possible  
because the T-duality action of the string theory background 
survives  the decoupling of the gravitational degrees of freedom.
We showed that the construction can be performed in both the type IIA and type IIB cases, either 
starting from the $\mathcal{N}=(2,0)$ or from the $\mathcal{N}=(1,1)$ case.
We have also provided the relation with the other constructions proposed in the literature
based on the S-fold projection.

There are many further questions that are of interest.
For example it may be interesting to extend the analysis to the $\mathcal{N}=(1,0)$ 
LST classified in \cite{Bhardwaj:2015oru} and to connect the T-duality twist discussed here
to the $\mathcal{N}=2$ S-folds studied in \cite{Apruzzi:2020pmv}.
 Another possible extension of the analysis regards the study of the $D_n$ and the exceptional case.
 In these cases the S-duality group descends from the T-duality group in the same way at it does 
 in the $A_n$ case, and it  implies that the results extend straightforwardly.
 A more difficult problem regards the non-simply laced cases. In these cases the structure of the 
 S-duality fixed point is different and the matter fields transforms under the S-duality group.

\section*{Acknowledgments}
We are grateful to Domenico Orlando, Marco Fazzi, Prarit Agarwal and Susanne Reffert. 
This work has been supported in part by Italian Ministero
dell'Istruzione, Universit\`a e Ricerca (MIUR), in part by Istituto Nazionale di Fisica
Nucleare (INFN) through the GSS research project and in part by MIUR-PRIN contract 2017CC72MK-003.

\bibliographystyle{ytphys}
\bibliography{Letteratura}

\providecommand{\href}[2]{#2}\begingroup\raggedright\begin{thebibliography}{10}

\bibitem{Gaiotto}
D.~{Gaiotto}, ``{N = 2 dualities},''
  \href{http://dx.doi.org/10.1007/JHEP08(2012)034}{{\em Journal of High Energy
  Physics} {\bfseries 2012} (Aug, 2012) 34},
  \href{http://arxiv.org/abs/0904.2715}{{\ttfamily arXiv:0904.2715 [hep-th]}}.

\bibitem{Weinberg:2000cr}
S.~Weinberg, {\em {The quantum theory of fields. Vol. 3: Supersymmetry}}.
\newblock Cambridge University Press,
2013.
\newblock

\bibitem{Garcia-Etxebarria:2015wns}
I.~García-Etxebarria and D.~Regalado, ``{$ \mathcal{N}=3 $ four dimensional
  field theories},'' \href{http://dx.doi.org/10.1007/JHEP03(2016)083}{{\em
  JHEP} {\bfseries 03} (2016) 083},
\href{http://arxiv.org/abs/1512.06434}{{\ttfamily arXiv:1512.06434 [hep-th]}}.

\bibitem{Aharony:2016kai}
O.~Aharony and Y.~Tachikawa, ``{S-folds and 4d N=3 superconformal field
  theories},'' \href{http://dx.doi.org/10.1007/JHEP06(2016)044}{{\em JHEP}
  {\bfseries 06} (2016) 044},
\href{http://arxiv.org/abs/1602.08638}{{\ttfamily arXiv:1602.08638 [hep-th]}}.

\bibitem{Garcia-Etxebarria:2016erx}
I.~García-Etxebarria and D.~Regalado, ``{Exceptional $ \mathcal{N}=3 $
  theories},'' \href{http://dx.doi.org/10.1007/JHEP12(2017)042}{{\em JHEP}
  {\bfseries 12} (2017) 042},
\href{http://arxiv.org/abs/1611.05769}{{\ttfamily arXiv:1611.05769 [hep-th]}}.

\bibitem{Bourton:2018jwb}
T.~Bourton, A.~Pini, and E.~Pomoni, ``{4d $\mathcal{N}=3$ indices via discrete
  gauging},'' \href{http://dx.doi.org/10.1007/JHEP10(2018)131}{{\em JHEP}
  {\bfseries 10} (2018) 131},
\href{http://arxiv.org/abs/1804.05396}{{\ttfamily arXiv:1804.05396 [hep-th]}}.

\bibitem{Argyres:2019ngz}
P.~C. Argyres, A.~Bourget, and M.~Martone, ``{Classification of all
  $\mathcal{N}\geq 3$ moduli space orbifold geometries at rank 2},''
\href{http://arxiv.org/abs/1904.10969}{{\ttfamily arXiv:1904.10969 [hep-th]}}.

\bibitem{Argyres:2019yyb}
P.~C. Argyres, A.~Bourget, and M.~Martone, ``{On the moduli spaces of 4d
  $\mathcal{N} = 3$ SCFTs I: triple special Kähler structure},''
\href{http://arxiv.org/abs/1912.04926}{{\ttfamily arXiv:1912.04926 [hep-th]}}.

\bibitem{Nishinaka:2016hbw}
T.~Nishinaka and Y.~Tachikawa, ``{On 4d rank-one $ \mathcal{N}=3 $
  superconformal field theories},''
  \href{http://dx.doi.org/10.1007/JHEP09(2016)116}{{\em JHEP} {\bfseries 09}
  (2016) 116},
\href{http://arxiv.org/abs/1602.01503}{{\ttfamily arXiv:1602.01503 [hep-th]}}.

\bibitem{Imamura:2016abe}
Y.~Imamura and S.~Yokoyama, ``{Superconformal index of ${ \mathcal N }=3$
  orientifold theories},''
  \href{http://dx.doi.org/10.1088/1751-8113/49/43/435401}{{\em J. Phys.}
  {\bfseries A49} no.~43, (2016) 435401},
\href{http://arxiv.org/abs/1603.00851}{{\ttfamily arXiv:1603.00851 [hep-th]}}.

\bibitem{Imamura:2016udl}
Y.~Imamura, H.~Kato, and D.~Yokoyama, ``{Supersymmetry Enhancement and
  Junctions in S-folds},''
  \href{http://dx.doi.org/10.1007/JHEP10(2016)150}{{\em JHEP} {\bfseries 10}
  (2016) 150},
\href{http://arxiv.org/abs/1606.07186}{{\ttfamily arXiv:1606.07186 [hep-th]}}.

\bibitem{Agarwal:2016rvx}
P.~Agarwal and A.~Amariti, ``{Notes on S-folds and $ \mathcal{N} $ = 3
  theories},'' \href{http://dx.doi.org/10.1007/JHEP09(2016)032}{{\em JHEP}
  {\bfseries 09} (2016) 032},
\href{http://arxiv.org/abs/1607.00313}{{\ttfamily arXiv:1607.00313 [hep-th]}}.

\bibitem{Garcia-Etxebarria:2017ffg}
I.~García-Etxebarria and D.~Regalado, ``{$N=3$ four dimensional field
  theories},'' \href{http://arxiv.org/abs/1708.03906}{{\ttfamily
  arXiv:1708.03906 [hep-th]}}.
[PoSCORFU2016,101(2017)].

\bibitem{vanMuiden:2017qsh}
J.~van Muiden and A.~Van~Proeyen, ``{The $ \mathcal{N} $ = 3 Weyl multiplet in
  four dimensions},'' \href{http://dx.doi.org/10.1007/JHEP01(2019)167}{{\em
  JHEP} {\bfseries 01} (2019) 167},
\href{http://arxiv.org/abs/1702.06442}{{\ttfamily arXiv:1702.06442 [hep-th]}}.

\bibitem{Cornagliotto:2017dup}
M.~Cornagliotto, M.~Lemos, and V.~Schomerus, ``{Long Multiplet Bootstrap},''
  \href{http://dx.doi.org/10.1007/JHEP10(2017)119}{{\em JHEP} {\bfseries 10}
  (2017) 119},
\href{http://arxiv.org/abs/1702.05101}{{\ttfamily arXiv:1702.05101 [hep-th]}}.

\bibitem{Borsten:2018jjm}
L.~Borsten, M.~J. Duff, and A.~Marrani, ``{Twin conformal field theories},''
  \href{http://dx.doi.org/10.1007/JHEP03(2019)112}{{\em JHEP} {\bfseries 03}
  (2019) 112},
\href{http://arxiv.org/abs/1812.11130}{{\ttfamily arXiv:1812.11130 [hep-th]}}.

\bibitem{Arai:2018utu}
R.~Arai, S.~Fujiwara, and Y.~Imamura, ``{BPS Partition Functions for
  S-folds},'' \href{http://dx.doi.org/10.1007/JHEP03(2019)172}{{\em JHEP}
  {\bfseries 03} (2019) 172},
\href{http://arxiv.org/abs/1901.00023}{{\ttfamily arXiv:1901.00023 [hep-th]}}.

\bibitem{Dabholkar:2002sy}
A.~Dabholkar and C.~Hull, ``{Duality twists, orbifolds, and fluxes},''
  \href{http://dx.doi.org/10.1088/1126-6708/2003/09/054}{{\em JHEP} {\bfseries
  09} (2003) 054},
\href{http://arxiv.org/abs/hep-th/0210209}{{\ttfamily arXiv:hep-th/0210209
  [hep-th]}}.

\bibitem{Losev:1997hx}
A.~Losev, G.~W. Moore, and S.~L. Shatashvili, ``{M \& m's},''
  \href{http://dx.doi.org/10.1016/S0550-3213(98)00262-4}{{\em Nucl. Phys.}
  {\bfseries B522} (1998) 105--124},
\href{http://arxiv.org/abs/hep-th/9707250}{{\ttfamily arXiv:hep-th/9707250
  [hep-th]}}.

\bibitem{Seiberg:1997zk}
N.~Seiberg, ``{New theories in six-dimensions and matrix description of M
  theory on T**5 and T**5 / Z(2)},''
  \href{http://dx.doi.org/10.1016/S0370-2693(97)00805-8}{{\em Phys. Lett.}
  {\bfseries B408} (1997) 98--104},
\href{http://arxiv.org/abs/hep-th/9705221}{{\ttfamily arXiv:hep-th/9705221
  [hep-th]}}.

\bibitem{Berkooz:1997cq}
M.~Berkooz, M.~Rozali, and N.~Seiberg, ``{Matrix description of M theory on
  T**4 and T**5},'' \href{http://dx.doi.org/10.1016/S0370-2693(97)00800-9}{{\em
  Phys. Lett.} {\bfseries B408} (1997) 105--110},
\href{http://arxiv.org/abs/hep-th/9704089}{{\ttfamily arXiv:hep-th/9704089
  [hep-th]}}.

\bibitem{Aharony:1999ks}
O.~Aharony, ``{A Brief review of 'little string theories'},''
  \href{http://dx.doi.org/10.1088/0264-9381/17/5/302}{{\em Class. Quant. Grav.}
  {\bfseries 17} (2000) 929--938},
\href{http://arxiv.org/abs/hep-th/9911147}{{\ttfamily arXiv:hep-th/9911147
  [hep-th]}}.

\bibitem{Kutasov:2001uf}
D.~Kutasov, ``{Introduction to little string theory},''
{\em ICTP Lect. Notes Ser.} {\bfseries 7} (2002) 165--209.

\bibitem{Dhokarh:2008ki}
D.~Dhokarh, S.~S. Haque, and A.~Hashimoto, ``{Melvin Twists of global AdS(5) x
  S(5) and their Non-Commutative Field Theory Dual},''
  \href{http://dx.doi.org/10.1088/1126-6708/2008/08/084}{{\em JHEP} {\bfseries
  08} (2008) 084},
\href{http://arxiv.org/abs/0801.3812}{{\ttfamily arXiv:0801.3812 [hep-th]}}.

\bibitem{V}
C.~{Vafa}, ``{Geometric Origin of Montonen-Olive Duality},'' {\em arXiv
  e-prints} (Jul, 1997) hep--th/9707131,
  \href{http://arxiv.org/abs/hep-th/9707131}{{\ttfamily arXiv:hep-th/9707131
  [hep-th]}}.

\bibitem{AKS}
P.~C. {Argyres}, A.~{Kapustin}, and N.~{Seiberg}, ``{On S-duality for
  non-simply-laced gauge groups},''
  \href{http://dx.doi.org/10.1088/1126-6708/2006/06/043}{{\em Journal of High
  Energy Physics} {\bfseries 2006} no.~6, (Jun, 2006) 043},
  \href{http://arxiv.org/abs/hep-th/0603048}{{\ttfamily arXiv:hep-th/0603048
  [hep-th]}}.

\bibitem{BUSCHER}
T.~Buscher, ``A symmetry of the string background field equations,''
  \href{http://www.sciencedirect.com/science/article/pii/0370269387907696}{{\em
  Physics Letters B} {\bfseries 194} no.~1, (1987) 59 -- 62}.

\bibitem{KW}
A.~{Kapustin} and E.~{Witten}, ``{Electric-magnetic duality and the geometric
  Langlands program},''
  \href{http://dx.doi.org/10.4310/CNTP.2007.v1.n1.a1}{{\em Communications in
  Number Theory and Physics} {\bfseries 1} (Jan, 2007) 1--236},
  \href{http://arxiv.org/abs/hep-th/0604151}{{\ttfamily arXiv:hep-th/0604151
  [hep-th]}}.

\bibitem{Bianchi:2011qh}
M.~Bianchi, A.~Collinucci, and L.~Martucci, ``{Magnetized E3-brane instantons
  in F-theory},'' \href{http://dx.doi.org/10.1007/JHEP12(2011)045}{{\em JHEP}
  {\bfseries 12} (2011) 045},
\href{http://arxiv.org/abs/1107.3732}{{\ttfamily arXiv:1107.3732 [hep-th]}}.

\bibitem{I}
K.~{Intriligator} and W.~{Skiba}, ``{Bonus symmetry and the operator product
  expansion of N = 4 super-Yang-Mills},''
  \href{http://dx.doi.org/10.1016/S0550-3213(99)00430-7}{{\em Nuclear Physics
  B} {\bfseries 559} no.~1, (Oct, 1999) 165--183},
  \href{http://arxiv.org/abs/hep-th/9905020}{{\ttfamily arXiv:hep-th/9905020
  [hep-th]}}.

\bibitem{Bhardwaj:2015oru}
L.~Bhardwaj, M.~Del~Zotto, J.~J. Heckman, D.~R. Morrison, T.~Rudelius, and
  C.~Vafa, ``{F-theory and the Classification of Little Strings},''
  \href{http://dx.doi.org/10.1103/PhysRevD.100.029901,
  10.1103/PhysRevD.93.086002}{{\em Phys. Rev.} {\bfseries D93} no.~8, (2016)
  086002}, \href{http://arxiv.org/abs/1511.05565}{{\ttfamily arXiv:1511.05565
  [hep-th]}}.
[Erratum: Phys. Rev.D100,no.2,029901(2019)].

\bibitem{Apruzzi:2020pmv}
F.~Apruzzi, S.~Giacomelli, and S.~Schäfer-Nameki, ``{4d $\mathcal{N}=2$
  S-folds},''
\href{http://arxiv.org/abs/2001.00533}{{\ttfamily arXiv:2001.00533 [hep-th]}}.

\end{thebibliography}\endgroup


\end{document}